\documentclass[twocolumn,showpacs,preprintnumbers,amsmath,amssymb]{revtex4}
\usepackage{graphicx}
\usepackage{dcolumn}
\usepackage{bm}
\usepackage{epsfig}
\usepackage{dcolumn}
\usepackage{bm}
\def\avgNp{${\rm \langle N_{part} \rangle }$}
\def\avgN2p{${\rm \langle N_{part}/2 \rangle }$}
\def\avgNpt{${\rm \langle N_{part} \rangle }$}
\def\avgNpAu{${\rm \langle N^{Au}_{part} \rangle}$}
\def\avgNpd{${\rm \langle N^{d}_{part} \rangle}$}
\def\avgNc{${\rm \langle N_{coll} \rangle }$}
\def\avgNch{ ${\rm N^{ch}_{ | \eta | \le 5.4}  }$}
\def\avgNEm{${\rm \langle N_{part}(pEm) \rangle}$}
\def\avgNdAu{${\rm \langle N_{part}(dAu) \rangle}$}
\def\avgNchs{${\rm N^{ch}_{Tot} }$}

\def\AuAu{${\rm Au}+{\rm Au}$}

\def\snn{${ \sqrt{s_{_{NN}}}~ = \rm {200~GeV} }$}
\def\sn{$\sqrt{s_{_{NN}}}$}

\def\dAu{${\rm d}+{\rm Au}$}

\def\p{${\rm p(\bar{p}) + p}$}

\def\dnch{${\rm dN_{ch}/d{\rm \eta} }$}

\begin{document}
\title{Scaling of Charged Particle Production in d+Au Collisions at~\snn}
\author{
B.B.Back$^1$,
M.D.Baker$^2$,
M.Ballintijn$^4$,
D.S.Barton$^2$,
B.Becker$^2$,
R.R.Betts$^6$,
A.A.Bickley$^7$,
R.Bindel$^7$,
W.Busza$^4$,
A.Carroll$^2$,
M.P.Decowski$^4$,
E.Garc\'{\i}a$^6$,
T.Gburek$^3$,
N.George$^2$,
K.Gulbrandsen$^4$,
S.Gushue$^2$,
C.Halliwell$^6$,
J.Hamblen$^8$,
A.S.Harrington$^8$,
C.Henderson$^4$,
D.J.Hofman$^6$,
R.S.Hollis$^6$,
R.Ho\l y\'{n}ski$^3$,
B.Holzman$^2$,
A.Iordanova$^6$,
E.Johnson$^8$,
J.L.Kane$^4$,
N.Khan$^8$,
P.Kulinich$^4$,
C.M.Kuo$^5$,
J.W.Lee$^4$,
W.T.Lin$^5$,
S.Manly$^8$,
A.C.Mignerey$^7$,
R.Nouicer$^{2,6}$,
A.Olszewski$^3$,
R.Pak$^2$,
I.C.Park$^8$,
H.Pernegger$^4$,
C.Reed$^4$,
C.Roland$^4$,
G.Roland$^4$,
J.Sagerer$^6$,
P.Sarin$^4$,
I.Sedykh$^2$,
W.Skulski$^8$,
C.E.Smith$^6$,
P.Steinberg$^2$,
G.S.F.Stephans$^4$,
A.Sukhanov$^2$,
M.B.Tonjes$^7$,
A.Trzupek$^3$,
C.Vale$^4$,
G.J.van~Nieuwenhuizen$^4$,
R.Verdier$^4$,
G.I.Veres$^4$,
F.L.H.Wolfs$^8$,
B.Wosiek$^3$,
K.Wo\'{z}niak$^3$,
B.Wys\l ouch$^4$,
J.Zhang$^4$\\
\vspace{3mm}
\small
$^1$~Argonne National Laboratory, Argonne, IL 60439-4843, USA\\
$^2$~Brookhaven National Laboratory, Upton, NY 11973-5000, USA\\
$^3$~Institute of Nuclear Physics PAN, Krak\'{o}w, Poland\\
$^4$~Massachusetts Institute of Technology, Cambridge, MA 02139-4307, USA\\
$^5$~National Central University, Chung-Li, Taiwan\\
$^6$~University of Illinois at Chicago, Chicago, IL 60607-7059, USA\\
$^7$~University of Maryland, College Park, MD 20742, USA\\
$^8$~University of Rochester, Rochester, NY 14627, USA\\
}
\begin{abstract}
The measured pseudorapidity distributions of primary charged particles
over a wide pseudorapidity range of ${\rm | \eta | \le 5.4 }$
and integrated charged particle multiplicities in\ \dAu\ collisions
at\ \snn\ are presented as a function of collision centrality.
The longitudinal features of\ \dAu\ collisions at\ \snn\ are found to be
very similar to those seen in p + A collisions at lower energies. The
total multiplicity of charged particles is found to scale with the total number
of participants according to ${\rm N^{dAu}_{ch} = \frac{1}{2}}$\avgNp
${\rm N^{pp}_{ch}}$ and the energy dependence of the density of
charged particles produced in the fragmentation
region exhibits extensive longitudinal scaling. 
\end{abstract}
\maketitle
Charged particle multiplicities have been studied extensively in high energy
collisions because of the intrinsic interest in understanding 
the production mechanism. More recent interest comes in the context
of searching for and studying new forms of matter that are expected to
be created in heavy ion collisions at ultrarelativistic energies. A key quantity
that contains information about the longitudinal aspects of the 
multiparticle production process, and that has provided 
valuable input for discriminating between phenomenological models
in the past \cite{ref1}, 
is the rapidity distribution of identified particles. When particle
identification is not available,
the almost equivalent pseudorapidity distribution of charged
particles suffices. For this reason, such distributions have been
studied in detail in hadron + proton \cite{ref2}, hadron + nucleus
\cite{ref3}, and nucleus + nucleus \cite{Sch1} collisions.

Since the first collisions were achieved at the Relativistic Heavy Ion Collider
(RHIC), the PHOBOS collaboration has obtained extensive 
information on pseudorapidity distributions of charged particles
produced in\ \AuAu\ collisions for nucleon-nucleon center-of-mass energies (\sn) between 19.6 and 200
GeV and over a large range of collision geometries. We have observed, for example, that the    
particle density in the midrapidity region changes smoothly as a
function of\ \sn\ \cite{Bac1} and that the
total charged particle production scales linearly with the number of
participants \cite{pet}. Further detailed observations of the shape of
the pseudorapidity distribution show a scaling according 
to the ``limiting fragmentation'' hypothesis \cite{Bac3}.
The study of a simpler system such as\ \dAu\ at the same energy
as\ \AuAu\ is essential to gain insight into which aspects of the data follow from the initial
interacting states or general properties of the hadronic production
process, and which are the consequence of the very different
environments created in\ \dAu\ and\ \AuAu\ collisions. 

In this Letter we present the results of detailed measurements of the
pseudorapidity distributions of primary charged particles,\ \dnch, as a function 
of collision centrality in\ \dAu\ collisions at\
\snn\ \cite{beam}
over a wide
pseudorapidity range of ${\rm | \eta | \le }$ 5.4. 
The pseudorapidity, ${\rm \eta }$, is defined as
${\rm \eta = -ln[tan(\theta/2)]}$, where ${\rm \theta }$
is the emission angle relative to the direction of the deuteron beam.
The results for\ \dAu\ collisions are compared to\ \AuAu\ collisions and inelastic
\p\ collisions at\ \snn\ as well as to
p($\pi^{+}$, K$^{+}$) + nucleus collisions at lower energies.

The data were obtained with the multiplicity array of
the PHOBOS detector \cite{BackNim} at RHIC. 
The array consists of an octagonal barrel of silicon detectors, the
Octagon, surrounding the interaction region in an approximately cylindrical geometry 
covering ${\rm | \eta | \le 3.2}$. This array is augmented by
two sets of three annular silicon counter arrays, the Rings, along the
beam pipe far forward and backward of the interaction
region (${\rm 3.0 < | \eta | < 5.4}$). This
array is identical to that used in our study
of\ \AuAu\ collisions \cite{Bac3}. The setup also includes two sets of 16
scintillator counters (${\rm 3.0 < | \eta | < 4.5}$)
which were used in the primary event trigger and in the offline event
selection. 

The centrality determination was based on the observed total energy deposited
in the Ring counters, E$_{\rm Ring}$, which is proportional to the number of
charged particles hitting these detectors. The choice of this
centrality measure was based on extensive studies utilizing both data
and Monte Carlo (MC) simulations that sought to minimize effects of
auto-correlations on the final dN$_{\rm ch}$/d$\eta$ result. These effects can
be significant when using other centrality measures \cite{dAumin},
resulting in enhancements (suppressions) in the reconstructed
midrapidity yields of up to $\sim$ 30\% for central (peripheral) collisions.
The MC simulations used in the study included both HIJING
\cite{HIJING} and AMPT \cite{AMPT} event
generators coupled to a full GEANT \cite{GEANT} simulation of the PHOBOS detector.
\begin{figure}[t] 
\hspace*{-0.5cm}\epsfig{file=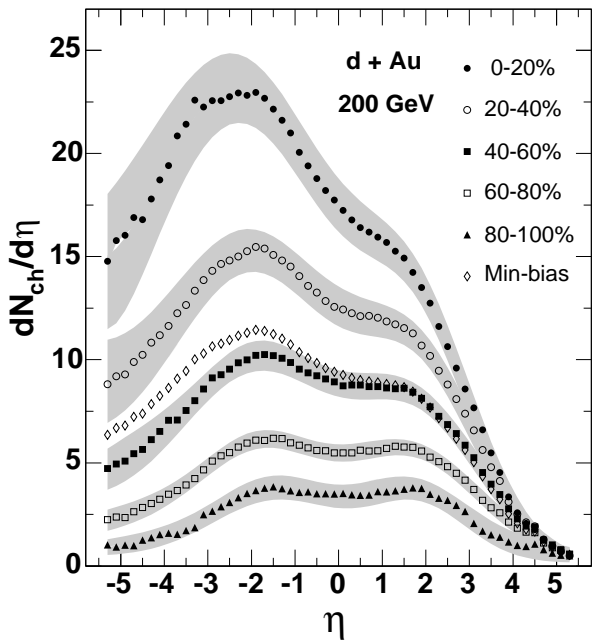,width=8cm}
\vspace*{-0.6cm}\caption{Measured pseudorapidity distributions of
charged particles from\ \dAu\ collisions at~\snn~as a
function of collision centrality. 
Shaded bands represent 90\% confidence level systematic errors and the 
statistical errors are negligible.
The minimum-bias distribution is shown as open
diamonds~\cite{dAumin}.}
\label{fig:fig1}
\end{figure}

Four additional centrality measures, discussed in Ref. \cite{dAumin},
were created in order to study the detailed effects of auto-correlation biases.
Ratios of the reconstructed dN$_{\rm ch}$/d$\eta$ distributions obtained from the five centrality measures
for data and, independently, for the MC simulations were found to
agree, giving confidence in the entire methodology. This information, together
with knowledge of the unbiased MC simulated ``truth'' distributions, provided a
clear choice of the centrality measure based on the Ring detectors as that
which yielded the least bias on the measurement.
It is important to note that this study only provided guidance with
respect to the choice of E$_{\rm Ring}$ for the
experimental centrality measure, and the final experimental dN$_{\rm
ch}$/d$\eta$ results do not rely in any way on the detailed shape of
the\ \dnch\ distributions from the MC simulations.

The multiplicity signals of E$_{\rm Ring}$ were divided into five centrality
bins, where each bin contained 20\% of the total cross section. For this to be
done correctly, the trigger and vertexing efficiency had to be determined for
each bin.  Knowledge of the efficiency as a function of multiplicity allowed
for the correct centrality bin determination in data as well as the extraction
of the corresponding efficiency-averaged number of participants. 
A comparison of the data and the MC simulations yielded an overall efficiency of $\sim$
83\%.

Results of the Glauber calculations implemented in the MC were
used to estimate the average total number of nucleon participants, \avgNp,
the number of participants in the incident gold, \avgNpAu, and the deuteron,
\avgNpd, nuclei, as well as the number of binary collisions, \avgNc, for each
centrality bin (see Table \ref{tab1}). 

The details of the analysis leading to the measurements of~\dnch~can
be found in Ref. \cite{Back2001}. 
The measured\ \dnch\ was corrected for particles which
were absorbed or produced in the surrounding material and for feed-down products from
weak decays of neutral strange particles. Uncertainties
in\ \dnch\ associated with these corrections range from 6\% in the
Octagon up to 28\% in the Rings. These uncertainties dominate the systematic errors.

\begin{figure}[t] 
\hspace*{-0.5cm}\epsfig{file=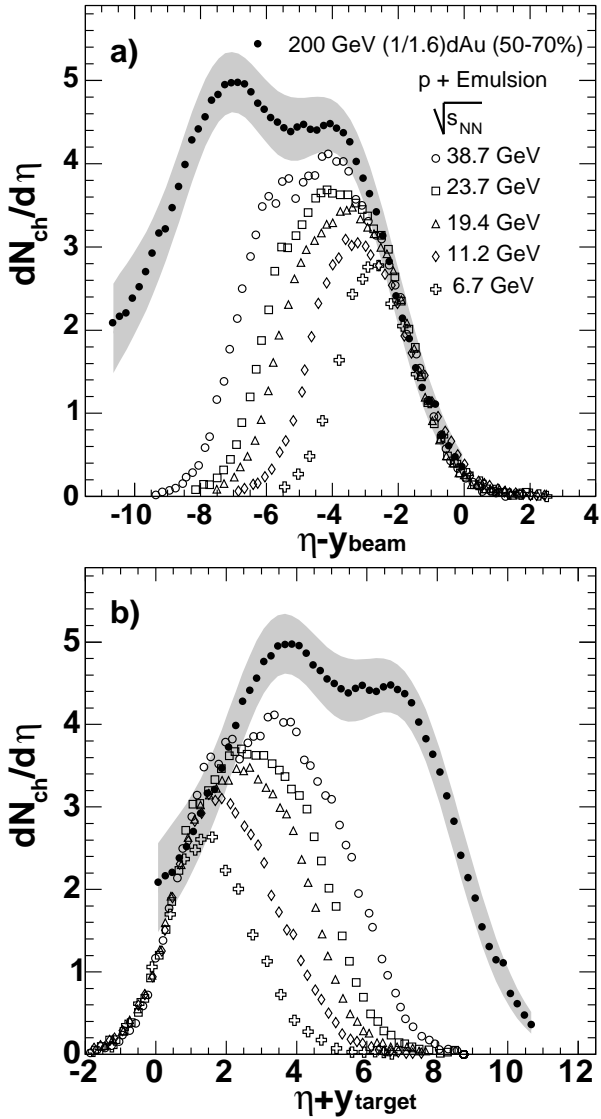,width=8cm}
\vspace*{-0.3cm}\caption{Panel a): Comparison of\ \dnch\ distributions 
for\ \dAu\ collisions at\ \snn\ to p + Em collisions 
(sum of shower and gray tracks) at five
energies~\cite{Sver87, Abd87}. 
The ${\rm \eta}$ measured in the center-of-mass 
system has been shifted to ${\rm \eta - y_{beam}}$ in
order to study the fragmentation regions in the deuteron/proton rest frame.  
Panel b): similar to panel~a) but shifted to ${\rm \eta + y_{target}}$ in
order to study the fragmentation regions in the gold/Emulsion rest frame.}
\label{fig:fig3}
\end{figure}

Figure~\ref{fig:fig1} shows the pseudorapidity distributions of
primary charged particles for\ \dAu\ collisions at\ \snn\ in five
centrality bins and for minimum-bias events. A detailed discussion of our
minimum-bias distribution can be found in Ref.~\cite{dAumin}. 
As a function of collision centrality, the
integrated charged particle multiplicity in the measured
region (${\rm | \eta | \le 5.4 }$) and the estimated total
charged particle multiplicity extrapolated to the unmeasured region using
guidance from the shifted p+nucleus data (see Fig. \ref{fig:fig3}) are presented in
Table \ref{tab1}.~The centrality bins 0-20\% and 80-100\% correspond to
the most central and the most peripheral collisions,
respectively. The pseudorapidity is measured in the
nucleon-nucleon center-of-mass frame; a negative pseudorapidity
corresponds to the gold nucleus direction. 
For the most central collisions, the mean ${\rm \eta }$ of the
distribution is found to be negative,  
reflecting the net longitudinal momentum
of the participants in the laboratory (NN) frame. 
For more peripheral collisions, the mean ${\rm \eta}$ tends to zero as
the distribution becomes more symmetric.
For measurements of\ \dAu\ in the nucleon-nucleon center-of-mass 
system the Jacobian between ${\rm dN_{ch}/dy}$ and\ \dnch\ naturally
produces the ``double-hump'' structure in\ \dnch\ even if there is no structure in ${\rm dN_{ch}/dy}$.
\begin{figure}[t] 
\hspace*{-0.5cm}\epsfig{file=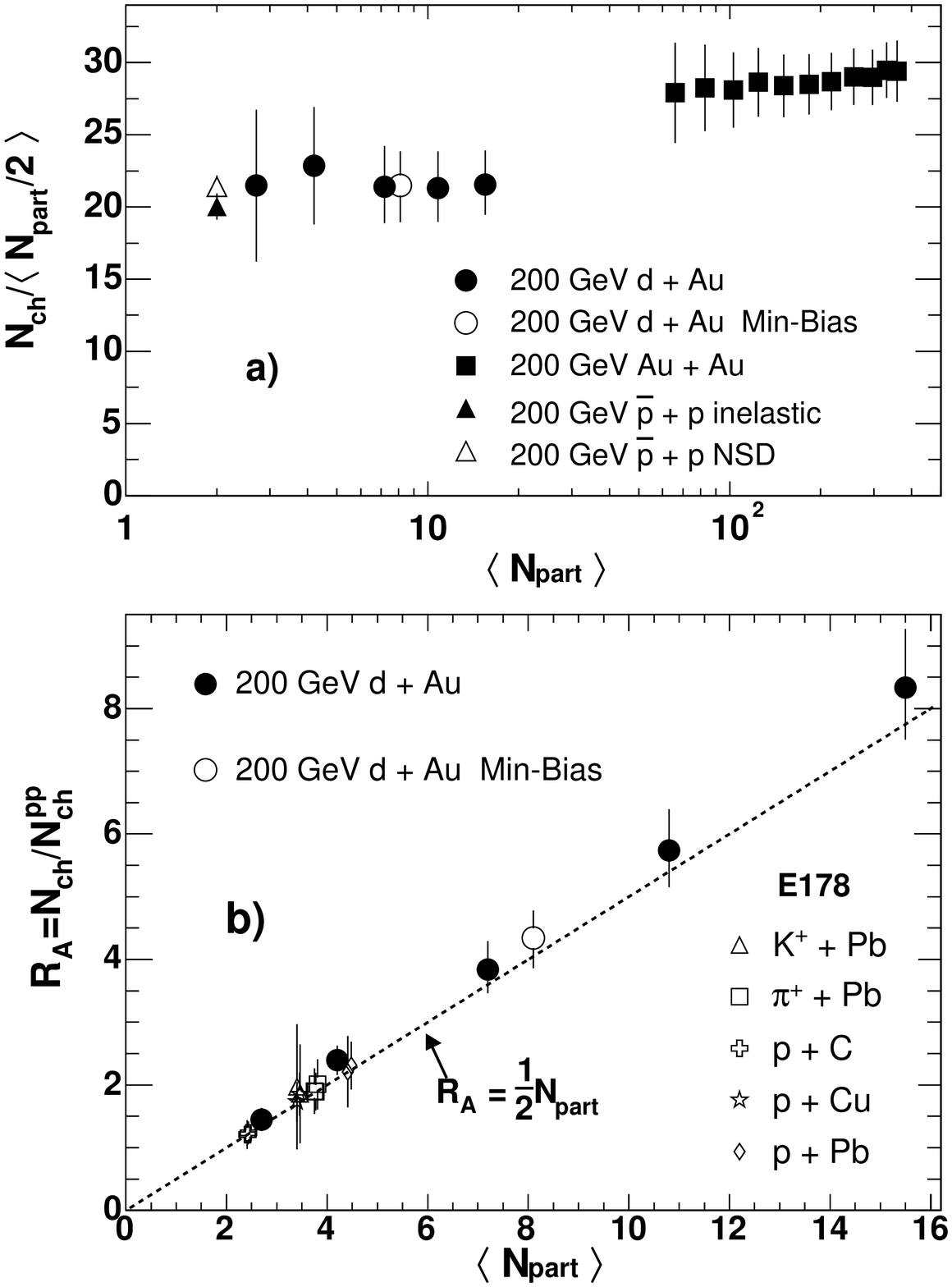,width=8cm}
\vspace*{-0.3cm}\caption{Panel a): Total integrated charged particle
multiplicity per participant pair in~\dAu,~\AuAu~\cite{pet},
~\p~inelastic \cite{nchine} and \p\ NSD \cite{nchnsd} collisions at the same
energy,~\snn.~Panel b): The ratio ${\rm R_{A}= {{N_{ch}}/{N^{pp}_{ch}}}}$, where ${\rm
N^{pp}_{ch}}$ is the total number of charged particles for inelastic\
\p\  collisions, as a
function of the total number of participant nucleons~\avgNp~for different collision
systems. The $\pi^{+}$+Pb, p+C, p+Cu and p+Pb collisions are for~\sn~= 19.4,
13.7 and 9.69 GeV, and the K$^{+}$+Pb for~\sn~= 13.7 and 9.69 GeV from
Ref.~\cite{Elias80}, respectively. 
For the\ \dAu\ and\ \AuAu\ data the error bars indicate the systematic errors
(90\% C.L.). 
The dashed line represents the linear
relation, ${\rm R_{A}={1 \over 2}}$\avgNp.}
\label{fig:fig4}
\end{figure}
\begin{figure}[t] 
\hspace*{-0.5cm}\epsfig{file=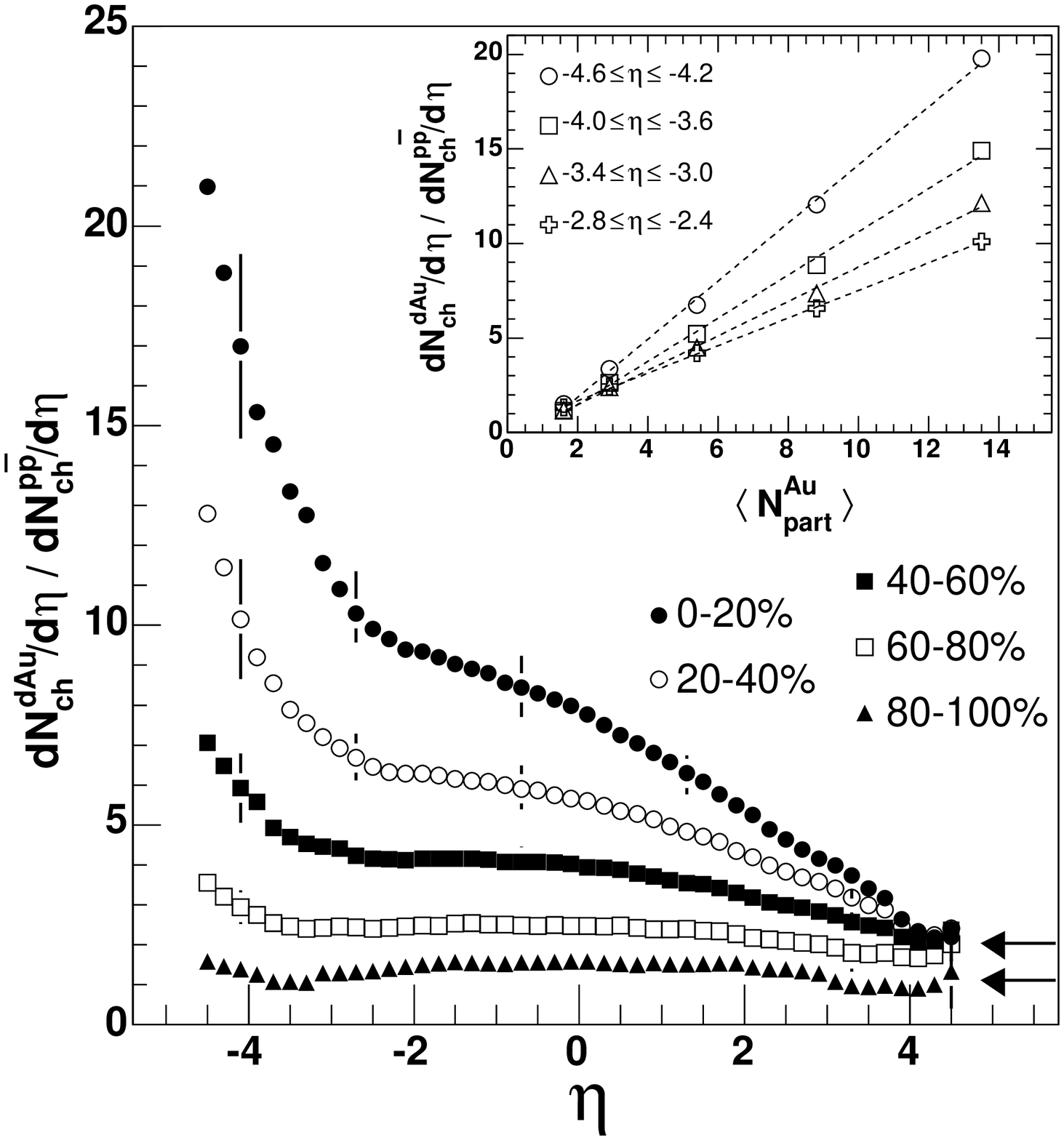,width=8cm}
\vspace*{-0.6cm}\caption{Centrality dependence of the\ \dnch\ ratio 
of~\dAu~collisions relative to that for inelastic~\p~collisions \cite{ua5dn} at the same energy. The arrows represent 
the \avgNpd~of the most central and the most peripheral collisions.
Typical systematic errors are shown for selected points (90\% C.L.). 
The inset figure shows the variation of the ratio as a function of
the number of gold participants, \avgNpAu, for four ${\rm
\eta}$ regions in the gold direction. The dashed lines represent
a linear fit to the data.}
\label{fig:fig2}
\end{figure}

Now, we compare our\ \dAu\ results with p + A data obtained at lower
energy, and discuss the energy and centrality dependence
of the data. 
Figure \ref{fig:fig3} compares\ \dnch\ distributions
of\ \dAu\ to p + Emulsion (Em) collisions at five
energies \cite{Sver87, Abd87}, in the effective rest frame of both the projectile
``beam'' (a) and target (b). 
For p + Em data the pseudorapidity distributions represent the sum of
shower and gray tracks. It should
be noted that ${\rm \eta}$ is measured in a different reference frame
for~\dAu\ and p~+~Em. This means that, compared to\ \dAu\ collisions, the p + Em pseudorapidity
distributions are suppressed by the Jacobian for ${\rm \eta +
y_{target} \sim 0}$.\
The 50-70\% centrality bin of\ \dAu\ collisions was selected in order to
match as well as possible 
${\rm {N^{Au}_{part}}/{N^{d}_{part}}}$ to 
${\rm {N^{Em}_{part}}/{N^{p}_{part}}}$ 
 where \avgNEm=3.4. 
The relative normalization of the\ \dnch\ for d + Au and p + Em collisions requires a
ratio of\ \avgNdAu/\avgNEm=1.6, such that the data correspond to the
same number of nucleons interacting with the nucleus. 
A remarkably good agreement (limiting fragmentation) is observed in the fragmentation
regions of the deuteron (gold) between d + Au and p + Em collisions at
different energies.  Furthermore, the overlap between the
fragmentation regions of the deuteron (gold) and proton (Em) extends
to lower ${\rm | \eta |}$ with increasing collision energy. In
Ref.~\cite{RNQM04} we reported a similar comparison but for more
central\ \dAu\ and p+Pb \cite{ref3} collisions.  This extensive
longitudinal scaling, also seen earlier in\ \p\ \cite{ref2}, p +
A~\cite{clv}, and A~+~A~\cite{Bac3} collisions, suggests that the
dominance of the two fragmentation regions is a common longitudinal
feature of all multiparticle production processes.

Figure \ref{fig:fig4}a shows the total integrated 
charged particle multiplicity per participant pair
for\ \dAu,\ \AuAu~\cite{pet} collisions, \p\
inelastic \cite{nchine} and\ \p\ NSD \cite{nchnsd} all at the same 
energy,\ \snn, as a function of the total number of participants. 
The results show that the 
total charged particle multiplicity scales linearly with\ \avgNp\ in both\ \dAu\ and\ \AuAu\ collisions. 
They also indicate that the transition between inelastic\ \p\ collisions 
and\ \AuAu\ collisions is not controlled simply by the number of
participants, as even very central\ \dAu\ multiplicity per participant
pair shows no signs of extrapolating to the\ \AuAu\ results. 
Not only do we find that the total charged particle production
in\ \dAu\ scales linearly with\ \avgNp, but also that the scaling is energy independent
and is the same in all hadron + nucleus collisions\ \cite{ref3}. This is 
evident from Figure\ \ref{fig:fig4}b 
where the ratio ${\rm R_{A}= { N_{ch} }/{N^{pp}_{ch}} }$ 
is plotted as a function of\ \avgNp\ for a large variety of systems and
energies. Here ${\rm N_{ch}}$ is the integrated total charged particle
multiplicity for d+ A, p + A\ \cite{Elias80} and ${\rm N^{pp}_{ch}}$ is for\ \p. 
It is this scaling, observed for the first time by Busza et
al.~\cite{wit2}, which led to the wounded
nucleon model of Bia\l as et al.\ \cite{Bial76}. 

Finally, in Figure \ref{fig:fig2} we compare for all centralities, 
the ratio of\ \dnch\ distributions observed in\ \dAu\ collisions relative
to that for inelastic\ \p\ \cite{ua5dn} at the same energy. 
The inset figure shows the ratio as function of
the number of gold participants, \avgNpAu, for four ${\rm
\eta}$ regions in the gold direction. 
The results are consistent with a picture in  which the production of 
particles with rapidity  near that of the incident deuteron (gold) is approximately
proportional to the number of deuteron (gold) participants. 
These results are consistent with lower
energy p + A data \cite{Elias80,Dem} and, 
with the quark-parton model of Brodsky et {\it al.} \cite{Bro77}.

In summary, we find that the longitudinal features of\ \dAu\ collisions
at\ \snn, as reflected by the centrality dependence of the
pseudorapidity distributions of charged particles, are very similar to
those seen in p~+~A collisions at energies lower by more than an order
of magnitude.~In particular, we find that in\ \dAu\ collisions the total
multiplicity of charged particles scales linearly with the total number of
participants, that the transition between the multiplicity per
participant in \dAu\ and\ \AuAu\ collisions 
is not controlled simply by the total number of participants, and that the
energy dependence of the density of charged particles
produced in the fragmentation regions exhibits extensive longitudinal
scaling. These results impose strong constraints on models
of multiparticle production. 

This work was partially supported by U.S. DOE grants
DE-AC02-98CH10886,
DE-FG02-93ER40802,
DE-FC02-94ER40818,  
DE-FG02-94ER40865,
DE-FG02-99ER41099, and
W-31-109-ENG-38, by U.S.
NSF grants 9603486, 
0072204,            
and 0245011,        
by Polish KBN grant 1-P03B-062-27(2004-2007), and
by NSC of Taiwan Contract NSC 89-2112-M-008-024.

\vspace*{-0.5cm}
\begin{table}[htpb]
\caption{\label{tab1}Estimated 
number of nucleon participants in the incoming gold (\avgNpAu) and deuteron
(\avgNpd) nuclei (\avgNpt = \avgNpd + \avgNpAu) and the number of collisions
(\avgNc) as a function of centrality in \dAu\ collisions. 
The integrated charged
particle multiplicity in the measured region (${\rm | \eta | \le
5.4 }$) and the estimated 
total charged particle multiplicity extrapolated to the unmeasured
region (see text) are listed. All errors are systematic.} 
\begin{ruledtabular}
\begin{tabular}{cccccc}
Cent. (\%) & \avgNpAu & \avgNpd & \avgNc & \avgNch & \avgNchs \\
\hline
 0--20  & 13.5 $\pm$ 1.0 &  2.0 $\pm$ 0.1 &  14.7
$\pm$ 0.9 & 157 $\pm$ 10 & 167${\rm ^{+14}_{-11}}$ \\ 
20--40& 8.9 $\pm$ 0.7 &  1.9 $\pm$ 0.1 & 9.8
$\pm$ 0.7 & 109 $\pm$ 7 & 115${\rm ^{+10}_{-8}}$ \\ 
40--60  & 5.4 $\pm$ 0.6 &  1.7 $\pm$ 0.2 & 5.9
$\pm$ 0.6 & 74 $\pm$ 5 & 77${\rm ^{+7}_{-5}}$ \\ 
60--80  & 2.9 $\pm$ 0.5 &  1.4 $\pm$ 0.2 & 3.1
$\pm$ 0.6 & 46 $\pm$ 3 & 48${\rm ^{+3}_{-3}}$ \\ 
80--100 & 1.6 $\pm$ 0.4 &  1.1 $\pm$ 0.2 & 1.7
$\pm$ 0.5 & 28 $\pm$ 3 & 29${\rm ^{+3}_{-3}}$ \\ 
Min-Bias &6.6 $\pm$ 0.5 &  1.7 $\pm$ 0.1 & 7.1
$\pm$ 0.5 & 82 $\pm$ 6 & 87${\rm ^{+7}_{-6}}$ \\ 
50-70 &3.9 $\pm$ 0.6 &  1.6 $\pm$ 0.2 & 4.2
$\pm$ 0.6 & 59 $\pm$ 4 & 62${\rm ^{+5}_{-4}}$ \\ 
\end{tabular}	
\end{ruledtabular}
\end{table}

\end{document}